\documentclass[a4paper,fleqn,usenatbib]{mnras}

\usepackage[T1]{fontenc}
\usepackage{ae,aecompl}

\usepackage{graphicx}   
\usepackage{amsmath}    
\usepackage{amssymb}    
\usepackage{color}

\newcommand{\Z}{\rm{Z}} 
\newcommand{\Zsun}{\rm{Z}_\odot} 
\newcommand{\Msun}{\rm{M}_\odot} 
\newcommand{\yr}{{\rm yr}} 
\newcommand{\pc}{\rm{pc}} 
\newcommand{\cSFR}{\dot{\rho}_{\rm{SFR}}(z)} 
\newcommand{\Zth}{Z_{\rm th}} 
\newcommand{\OH}{12 + \log (\textnormal{O}/\textnormal{H})}

\title[LGRB hosts in the Illustris simulation]{The metallicity and star formation activity of long gamma-ray burst hosts for z$<$3: insights from the Illustris simulation.}

\author[L. A. Bignone et al.]{
L. A. ~Bignone$^{1}$\thanks{E-mail:lbignone@iafe.uba.ar},
P. B. ~Tissera$^{1,2}$,
L. J. ~Pellizza$^{4}$
\\
$^{1}$Instituto de Astronom\'ia y F\'isica del Espacio (IAFE, CONICET-UBA), C.C. 67 Suc. 28, C1428ZAA Ciudad de Buenos Aires, Argentina.\\
$^{2}$Departamento de Ciencias F\'isicas, Universidad Andres Bello, Av. Republica 220, Santiago, Chile.\\
$^{4}$Instituto Argentino de Radioastronom\'ia (CCT-La Plata, CONICET; CICPBA), C.C. No. 5, 1894,Villa Elisa, Argentina.}

\date{Accepted XXX. Received YYY; in original form ZZZ}

\pubyear{2017}

\begin{document}
\label{firstpage}
\pagerange{\pageref{firstpage}--\pageref{lastpage}}
\maketitle

\begin{abstract}

We study the properties of long gamma-ray bursts (LGRBs) using a large scale
hydrodynamical cosmological simulation, the Illustris simulation. We determine
the LGRB host populations under different thresholds for the LGRB progenitor
metallicities, according to the collapsar model. We compare the simulated
sample of LGRBs hosts with recent, largely unbiased, host samples: BAT6 and
SHOALS. We find that at $z<1$ simulated hosts follow the mass-metallicity
relation and the fundamental metallicity relation simultaneously,
but with a paucity of high-metallicity hosts, in accordance with observations.
We also find a clear increment in the mean stellar mass of LGRB hosts and
their SFR with redshift up to $z<3$ on account of the metallicity dependence
of progenitors. We explore the possible origin of LGRBs in metal rich
galaxies, and find that the intrinsic metallicity dispersion in galaxies could
explain their presence. LGRB hosts present a tighter correlation between
galaxy metallicity and internal metallicity dispersion compared to normal star
forming galaxies. We find that the Illustris simulations favours the existence
of a metallicity threshold for LGRB progenitors in the range  $0.3 - 0.6 \,
\Zsun$.

\end{abstract}

\begin{keywords}

gamma-ray burst: general -- galaxies: abundances, evolution -- methods: numerical

\end{keywords}



\section{Introduction}

The study of the galaxy population hosting long gamma-ray Bursts (LGRBs)
provides a fundamental source of information to constrain the nature of
LGRB progenitors. For example, the link between LGRBs and the death of
massive stars \citep{hjorth_very_2003,hjorth_optically_2012} is supported by a
clear preference of LGRBs to be found in star-forming galaxies
\citep{le_floch_are_2003,savaglio_galaxy_2009}. As tracer of star formation,
LGRBs offer unique advantages since their intrinsic high luminosity makes them
detectable up to very high redshifts \citep[$z>6$,][]{tanvir_star_2012,basa_constraining_2012,salvaterra_high_2015}. This makes them promising tools to
study star formation in faint and distant galaxies
\citep[e.g.][]{kistler_unexpectedly_2008,robertson_connecting_2012,greiner_gamma-ray_2015}, a problem that would be otherwise difficult to
tackle using traditional star-forming galaxy surveys that can be affected by
redshift incompleteness, magnitude limits and dust extinction.

Evidence suggests that star formation is not the only factor regulating
the production of LGRBs. If that were the case, the LGRB population would
sample star-forming galaxies with a probability proportional to the galaxy
star formation rate (SFR). However, LGRB hosts show a preference for bluer and
less luminous galaxies \citep{le_floch_are_2003}, and crucially, with a lower
metallicity than typical star forming galaxies \citep{stanek_protecting_2006, modjaz_measured_2008,
savaglio_galaxy_2009,graham_metal_2013,jimenez_reconciling_2013}. The
existence of other factors which might influence the production of LGRBs is
still a matter of debate \citep{heuvel_are_2013}, but metallicity is  often
cited as a major regulator of the LGRB rate
\citep[e.g.][]{wolf_metallicity_2007,
kocevski_modeling_2009,trenti_luminosity_2015}.

Low metallicity is often predicted by core-collapse progenitor models
which require the retention of the angular momentum of the central core
to launch the jet in the first place  \citep{woosley_gamma-ray_1993,macfadyen_collapsars:_1999,izzard_formation_2004}. Single progenitor models predict very low metal progenitors in the range $\Z < 0.1-0.3 \,
\Zsun$
\citep{woosley_progenitor_2006}, while binary progenitor models favour a less
stringent metallicity dependence
\citep{izzard_formation_2004,fryer_binary_2005,podsiadlowski_explosive_2010}.
Hence, the metallicity of LGRB hosts  is an important observable to
distinguish between these two families of models.

Observationally, there has been significant improvements in the construction
of LGRB host samples. Early samples suffered from both heterogeneous
observations of galaxy properties and the probable presence of selection
biases \citep{le_floch_are_2003,fruchter_long_2006,savaglio_galaxy_2009,levesque_high-metallicity_2010,svensson_host_2010,mannucci_metallicity_2011,graham_metal_2013,perley_population_2013,hunt_new_2014}. Hosts samples
selected from optical afterglows, which are subjected to dust extinction,
tend to under-represent the more metal-rich and more massive hosts that are
typically related to heavily obscured or dark LGRBs \citep{fynbo_low-resolution_2009,perley_population_2013}.

Current samples are instead constructed from the X-ray LGRB emission, which is
largely unaffected by dust attenuation, and follow a carefully chosen set of
selection criteria. Among these, are the BAT6
\citep{salvaterra_complete_2012}, TOUGH \citep{hjorth_optically_2012} and
SHOALS \citep{perley_swift_2016} samples. Also redshift completeness has been
greatly improved, reaching values close to 90 percent for
the 69 objects in the TOUGH sample, $\geq 90$ percent for the 58 objects
in the BAT6 sample and 68 percent for the larger SHOALS sample composed
of 119 objects.

For these new host samples, detailed and homogeneous follow up observations
have allowed a clearer picture to emerge, at least, at low redshift (z $\leq
1.5)$. \citet{vergani_are_2015} found a strong deficiency of higher stellar
mass galaxies in the BAT6 sample at low redshift, as shown by their K-band
luminosity. These hosts were further studied by \citet{japelj_are_2016} who
analysed the detected nebular emission lines to measure the dust extinction,
SFR and nebular metallicity. They found that the high mass deficiency is
accompanied by a paucity of metal-rich LGRB hosts. Similar results were found
for the SHOALS sample by \citet{perley_swift_2016-1} who studied the irac 3.6 $\mu m$ luminosity which can be readily correlated with stellar mass. They
found that the host stellar mass increases with redshift, and that
such trend could be explained by a model in which the LGRB rate is
uniform with respect to metallicity below solar but drops by about an order
of magnitude in metal-richer galaxies. Similarly,
\citet{schulze_optically_2015} found that hosts in the TOUGH sample favour
lower UV luminosities, specially at low redshifts and that the UV brighter
hosts are located in the $1 < z < 3$ range. These results, reproduced by
independent and largely unbiased samples suggest that, at low redshift, the LGRBs
are indeed produced preferentially in low-metallicity environments.

Some of these studies predict higher metallicities thresholds
\citep[e.g.][]{perley_swift_2016-1} than the ones favoured by single
progenitor models. It is still unclear to what extent mean host metallicities are
representative of LGRB progenitor metallicities. It could be still possible
for low-metallicity star formation to occur in globally high-metallicity hosts either
due to metallicity gradients, accretion of poorly enrich
material or metallicity inhomogeneity in the interstellar medium (ISM) of
galaxies \citep{niino_revisiting_2011}.

The metallicity dependence of LGRB host galaxies have been studied using
different models and approximations, most of them based on the hypothesis that
the host metallicity is traced by the metallicity of the stellar progenitor or
vice-versa. For example, \citet{wolf_metallicity_2007} studied the efficiency
of producing LGRBs assuming that LGRB hosts followed the luminosity-metallicity relation of star-forming galaxies. They found that if a
metallicity cut-off existed for the formation of the LGRBs, it was in the
order of solar metallicity. Similarly, \citet{trenti_luminosity_2015}
predicted the luminosities, stellar masses, and metallicities of LGRB hosts
assuming that they followed the mass-metallicity relationship (MZR) derived by
\citet{maiolino_amaze._2008} and that LGRBs could be produced trough both a
metallicity dependent and independent channels.  They found a moderate
metallicity bias, where most LGRBs (80\%) are in very low metallicity environments
produced by collapsars and the remaining 20\% are produced by the
metallicity independent channel.

Alternatively,  theoretical approaches using both semi-analytic models of
galaxy formation
\citep{lapi_long_2008,campisi_properties_2009,chisari_host_2010} and
cosmological hydrodynamical simulations
\citep{nuza_host_2007,niino_revisiting_2011,salvaterra_high_2015}
provide a more consistent description of the chemical properties of
the local ISM and of  the global galaxy. In particular, considering
that  galaxy formation is a highly non-linear process, and
that the computation of LGRB host properties assuming a metallicity bias
depends strongly on a detailed knowledge of galaxy assembly,
cosmic star-formation and chemical evolution of the ISM, cosmological
hydrodynamical simulations
\citep{katz_dissipational_1991,navarro_simulations_1993,mosconi_chemical_2001,springel_cosmological_2003,scannapieco_feedback_2005,scannapieco_feedback_2006}
are specially  suitable for this task since they provide a
self-consistent description of these processes within a cosmological
context. 

Early studies using hydrodynamical simulation yielded results consistent with the existence of a
metallicity threshold for the triggering of LGRBs events
\citep[][]{courty_host_2004,nuza_host_2007}.
These simulations implemented simple subgrid physics and described
small cosmological volumes, which limited  galaxy statistics and made  difficult the study of rare, more massive, galaxies,
while lower resolutions made the results of low mass galaxies more
uncertain.
Current hydrodynamic simulations have dramatically
improved this situation thanks to larger volumes and more sophisticated
models. In particular, the cosmological hydrodynamical simulation
so-called Illustris
\citep{vogelsberger_introducing_2014,vogelsberger_properties_2014} attempts
to reproduce the galaxy population using state-of-the-art star formation,
supernovae and AGN feedback mechanisms. The Illustris simulation reproduces many observed
quantities, among them, the stellar mass distribution of galaxies, their
morphology, and gas content
\citep{vogelsberger_properties_2014,genel_introducing_2014}. The big advantage of the Illustris
simulation is its large cosmological volume, $\sim100^3$ Mpc$^3$. Hence, for the first time it is possible to construct a simulated LGRB
host galaxy sample covering different environments and allowing a more
suitable comparison with observations. Previous studies used small
volumes or resorted to semi-analytical techniques \citep[e.g.][]{courty_host_2004,nuza_host_2007,campisi_properties_2009,artale_chemical_2011,salvaterra_simulating_2013}.


In this work, for the first time, we propose to study the metallicity
dependant model using a cosmological significant LGRB galaxy sample selected
from the the state-of-the-art hydrodynamic cosmological Illustris simulation.
We compare the SFR, sSFR, metallicity and mass distributions of the LGRB hosts
predicted by the Illustris simulation to the complete and largely unbiased
host samples BAT6 and SHOALS. Furthermore, the nature of the hydrodynamical
simulation  allows us to study simultaneously the metallicity of the LGRB
hosts and the metallicity dependence that acts upon LGRB progenitors.
Therefore we also aim at studying the origin of high-metallicity host
galaxies.

The paper is organized as follows. Section 2 describes the
simulations and the LGRB progenitor model. In Section 3 the
properties of the LGRB host galaxies are analysed and compared to
observations.
Section 4  discusses the relation between the metallicity dispersion
of the galaxy hosts and the metallicity of the LGRB progenitor.
In Section 5 we describe the mass distribution of the simulated LGRB
hosts in a cosmological context. In Conclusions, the main results are summarized.

\section{Numerical models}

\subsection{Overview of the Illustris Simulation}

The Illustris project \citep{vogelsberger_introducing_2014,genel_introducing_2014}
consists of a series of large-scale hydrodynamical cosmological simulations with periodic
box 106.5 Mpc a side, computed using the quasi-Lagrangian \textsc{arepo} code
\citep{springel_e_2010}.  
The Illustris project adopted the following set of cosmological parameters:
$\Omega_m = 0.2726$, $\Omega_\Lambda = 0.7274$, $\Omega_b = 0.0456$, $\sigma_8
= 0.809$, $n_s = 0.963$ and $h = 0.704$, which are consistent with the
Wilkinson Microwave Anisotropy Probe \mbox{(WMAP)-9} measurements \citep{hinshaw_nine-year_2013}.

In our work we use  the highest resolution
simulation (Illustris-1, hereafter Illustris simulation). This
simulation has a dark mass resolution of $6.26 \times 10^6 \, \Msun$, and an
initial baryonic gas mass resolution of $1.26 \times 10^6 \, \Msun$. The
number of initial resolution elements for both baryonic and dark matter is
$1820^3$. The gravitational softening scale is $710 \, \pc$ at $z=0$, while
the smallest gas cells extend to $48 \, \pc$. At $z=0$ well-resolved galaxies (i.e. those with more than $500$ stellar particles) have minimum stellar masses in the range $10^{8-9} \Msun$, which is typical of LGRB hosts.

\begin{figure}
    \centering
    \includegraphics[width=1.0\columnwidth]{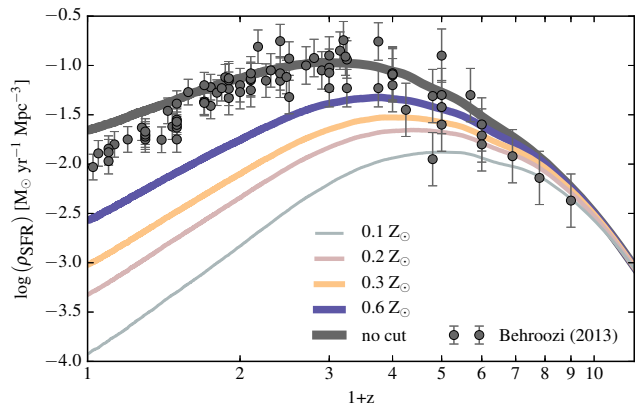}
    
    \caption{The cosmic star formation rate density for the Illustris
        simulation (thick grey line) and estimations using
                stellar populations with metallicity below a given  metallicity threshold:
        $0.1$ (green line), $0.2$ (brown line), $0.3$ (yellow line) and
                $0.6 ({\rm blue \, line}) \, \Zsun$). For comparison, the observations compiled by \citet{behroozi_average_2013} (circles) are included.}
    
    \label{fig:rho_sfr}
\end{figure}

The Illustris simulations follows the gravitational and dynamical
evolution of galaxies within a cosmological context, including  primordial and metal line radiative cooling with self-shielding
corrections, stellar evolution with associated mass loss and chemical
enrichment of the interstellar medium (ISM),  stellar wind feedback driven by
SNe,  super-massive black hole (SMBH) growth by accretion and mergers; and
active galactic nuclei (AGN) feedback with radio, quasar and radiative
modes \citep{vogelsberger_model_2013,marinacci_diffuse_2014,torrey_model_2014}.
The Illustris simulation is
able to reproduce reasonably well many of the key observable trends in the
local Universe, with some discrepancies related to the stellar ages of low
mass galaxies and the quenching of massive galaxies
\citep{vogelsberger_properties_2014}.

The dark matter halos were identified using the standard \mbox
{friends-of-friends} (FoF) algorithm \citep{davis_evolution_1985} with linking
length of 0.2 times the mean particle separation and a minimum number of 32
particles. Other particle types (stellar particles, gas cells, SMBH particles)
were assigned to the FoF group of the closest DM particle. Finally,
gravitationally bound substructures within the FoF groups were identified
using the \textsc{subfind} algorithm
\citep{springel_populating_2001,dolag_substructures_2009}. We refer to these
subhalos, when mentioning galaxies in the simulation. Each galaxy has a well
defined stellar  mass based on the particles which are bound to it. Throughout the rest
of the paper, we refer to this mass as the stellar mass of a galaxy.

Our sample of simulated galaxies is selected to have a minimum stellar mass
of $10^{8} \Msun$. From this galaxies we build up the LGBR host galaxy
sample which is the largest simulated sample of LGRB host galaxies hitherto
analysed in this kind of simulation. At $z=0$, they constitute 78,608 galaxies 
located in different environments in the $106.5$ Mpc a side cubic volume.

\subsection{LGRB population synthesis model}

If we assume that the stellar progenitors  of LGRBs are massive stars
with with a minimum mass
$M_{\rm{min}}$, then the LGBR rate should follow the star formation
rate in a perfectly unbiased way.
In this case the intrinsic cosmic LGRB rate can be related to the cosmic star formation rate density ($\cSFR$) as
\begin{equation}
    \Psi_{\rm{LGRB}}(z) = \frac{\int_{M_{\rm{min}}}^{100 \, \Msun} \psi(m) dm}{\int_{0.1 \, \Msun}^{100 \, \Msun} m \psi(m) dm} \dot{\rho}_{\rm{SFR}}(z),
    \label{eq:grb_sfr_frac}
\end{equation}

\noindent where $\psi$ is the initial mass function \citep[][in the case of
the Illustris simulation]{chabrier_galactic_2003} with lower and upper stellar
mass limits of $0.1 \, \Msun$ and $100 \, \Msun$, respectively. Within
this scenario, the production
of LGRBs is not significantly delayed with respect to the star formation events that created
their progenitor stars, since they are massive, short-lived objects.

Fig. \ref{fig:rho_sfr} shows the cosmic star formation rate density
for the Illustris simulation compared to observations
\citep{behroozi_average_2013}.  The predicted $\cSFR$ is in good agreement with
observations at intermediate and high redshifts but presents some tension at
lower redshifts where the simulated star formation is higher than expected. The
excess of star
formation appears to be related to insufficient AGN feedback in lower stellar-mass
galaxies \citep[see][for a more
detailed discussion]{vogelsberger_properties_2014,genel_introducing_2014} that contribute the most star formation at lower
redshift. Higher stellar-mass galaxies at low redshift are effectively quenched by AGN
feedback from central SMBHs.

A preference of LGRBs for low-metallicity progenitors can be easily included
by considering only the contribution to the $\cSFR$ from young stars with a
metallicity below a selected threshold ($Z_{\rm{th}}$). In
Fig~\ref{fig:rho_sfr} we show the  $\cSFR$  estimated by adopting  $Z_{\rm{th}}
= 0.1, 0.2, 0.3$ and $0.6 \, \Zsun$. As can be seen, the peak of the
relation shifts to higher redshift for lower metallicity
thresholds.
The intrinsic LGRB rate for a given $Z_{\rm{th}}$ is then
\begin{equation}
    \Psi_\text{LGRB}(z, Z_\text{th}) = \Sigma(z, Z_\text{th}) \Psi_\text{LGRB}(z),
    \label{eq:distr}
\end{equation}

\noindent  where $\Sigma(z,Z_{\rm th})$ is the fraction of newborn stellar
mass with metallicities below $Z_{\rm{th}}$. $\Sigma(z,Z_{\rm th})$ depends on
the baryonic physics  implemented in the simulations or analytical models.
In hydrodynamical simulations, such as Illustris, stellar populations are
represented by particles with an assigned mass, age and metallicity. Therefore, when
determining $\Sigma(z,Z_{\rm th})$ we take into account only individual stellar particles
with age $< 10$ Myr and metallicity below $Z_{\rm{th}}$.

\section{Simulated LGRB host galaxies}

Following previous works \citep{campisi_properties_2009, chisari_host_2010,artale_chemical_2011}, we assign each galaxy
a probability of being considered an LGRB host. This probability is then used as
a weight to compute the distribution of the physical properties of the
synthetic host galaxies sample such as
stellar mass,
SFR, metallicity, etc. These weighted distributions of observable synthetic parameters can be
compared to the corresponding observational
distributions. For a galaxy to be considered a host, at least one LGRB must be detected by a high-energy satellite. Then, the
galaxy itself must be detected. A significant advantage of the state-of-the-art LGRB galaxy samples, such as
BAT6, TOUGH and SHOALS, is that they select hosts exclusively based on the
properties of the X-ray LGRB emission and favourable observing
conditions. As consequence, biases related to the  detection of the host
galaxy are largely eliminated. This suggests to take the host probability $p_{\text{host}}$ proportional to the contribution of each galaxy to the observed LGRB rate. As all distribution will be computed at constant redshift, all cosmological related terms, as well as factors related to detector sensitivity, become constant. Therefore $p_{\text{host}}$ can be further simplified taking it proportional to the intrinsic number of LGRBs produced in a given galaxy (g),
\begin{equation}
N(g, \Zth) = m_*(g, \Zth) \frac{\int_{M_{\rm{min}}}^{100 \, \Msun} \psi(m) dm}{\int_{0.1 \, \Msun}^{100 \, \Msun} m \psi(m) dm},
\end{equation}

\noindent where $m_*(g, \Zth)$ is the total stellar mass of young stars (age $< 10$ Myr) with metallicity below $\Zth$.

\subsection{Mass-metallicity relation}
\label{sec:mzr}

\begin{figure*}
    \centering
    \includegraphics[width=1.0\textwidth]{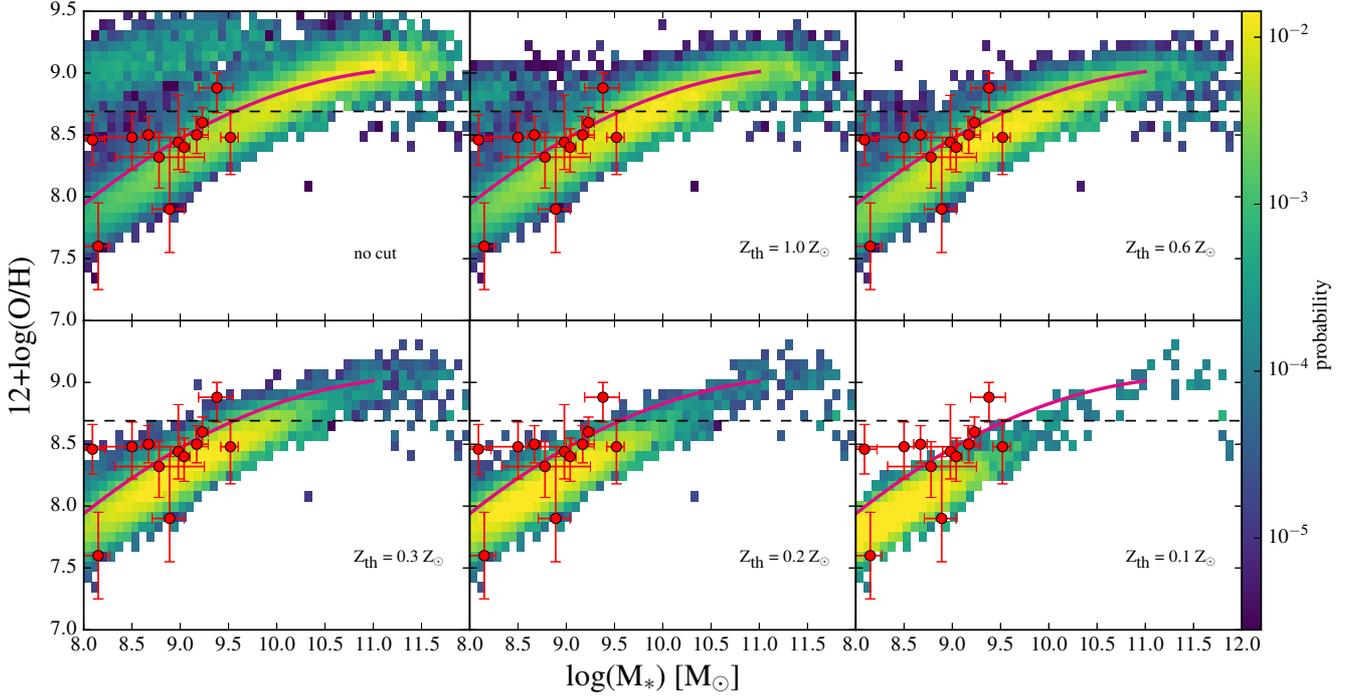}
    
    \caption{Mass-metallicity relation for host galaxies in the Illustris simulation at $z=0.7$. In each panel the bins are weighted by the probability of hosting an LGRB according to a different metallicity threshold. For comparison to observations we show the \citet{japelj_are_2016} MZR for the BAT6 sample in the M08 calibration (points). Also shown is the MZR of field galaxies using the parametrised analytical model of \citet{maiolino_amaze._2008} (solid lines)}

    \label{fig:mzr}
\end{figure*}

Considering that the metallicity is assumed to play a significant role
in the triggering of LGRBs, the MZR of the LGRB host galaxies in relation to
the whole galaxy population is an important aspect to analyse. In
order to perform direct comparisons to observations of galactic gas-phase
metallicities, we obtain from the simulation the mass-weighted average
metallicity of the gas cells bound to each galaxy, but restricted to cells
which are star forming. This places an emphasis on star-forming gas, mimicking
observational metallicity estimates from nebular emission lines. Particularly,
the Illustris project does not track individual chemical elements, rather,
they provide the metallicity as the ratio of the total gas mass of elements
heavier than He to the total galaxy gas mass. In order to compare our results
to observations, we convert the provided values to a metallicity expressed in
terms of the oxygen-to-hydrogen abundance ratio ($\OH$).  To do so a
primordial solar metal mass fraction value of $\Zsun = 0.0127$
\citep{wiersma_chemical_2009} and a solar oxygen abundance value of $\OH
= 8.69$ \citep{asplund_chemical_2009} are assumed.

In Fig.~\ref{fig:mzr} we compare the simulated MZR of the synthetic
LGRB host sample to the MZR of observed LGRB hosts obtained by \citet{japelj_are_2016} for the BAT6 sample, limited to $0.3 < z <1$. The mean redshift of this observational sample is $\sim 0.7$, which coincides with the redshift of the simulated hosts displayed in this figure. Here we show only the host metallicities in the
\citet{maiolino_amaze._2008} calibration, but analogous conclusions can be
extracted from the \citet{kobulnicky_metallicities_2004} calibration. 

Also, we compare
our results to the MZR of field galaxies using the parametrised analytical
model of \citet{maiolino_amaze._2008}. As pointed out by \citet{japelj_are_2016}, at sub-solar metallicities, observed LGRB hosts appear fairly consistent with the field MZR relation. In particular, there are no
signs of a systematic offsets towards values below the relation as was the case in
previous studies
\citep{mannucci_metallicity_2011,campisi_metallicity_2011}. This could
have been produced by 
incomplete samples
\citep[e.g.][]{levesque_high-metallicity_2010}. Instead, there is a
clear paucity 
of high stellar-mass hosts above the solar metallicity.

If no metallicity cut-off for LGRB progenitors is assumed, the MZR of simulated hosts appears
very different from observations. In that case, high metallicity
galaxies have the largest probability of hosting LGRBs, due to their
high SFRs. Only a strong metallicity threshold can
compensate for this effect. The lower panels of Fig~\ref{fig:mzr} shows the results of considering $\Zth = 0.3, 0.2$ and $0.1 \Zsun$. As can be seen, a significant shift towards lower metallicity hosts can be appreciated.

It is worth pointing out that the MZR of the Illustris simulation provides a
good match to the field star-forming galaxies MZR but with some differences,
as results evident in the first panel of Fig~\ref{fig:mzr}.  First, the simulated relation presents a stepper slope than observations, with a significant number of high stellar-mass galaxies appearing above the \citet{maiolino_amaze._2008} relation, while lower
stellar-mass ones appear below it. Secondly, a substantial
population of low stellar-mass, high-metallicity galaxies are found which do
not have an observational counterpart. These issues appear to be related to
the SN feedback model adopted in the Illustris simulation as discussed
in detail by  \citet{torrey_model_2014}. The origin of the steeper relation is
suggested to be due to metal low retention efficiency of low stellar-mass
galaxies due to galactic winds with large mass loading factor. But the
opposite seems to be the case for a certain subpopulation of low, stellar-mass
galaxies where metal retention is too high. Regardless of these issues, the
conclusion drawn in the previous paragraphs remains robust in the sense
that LGRB hosts appear to indeed follow the MZR intrinsic to the simulation
and that a metallicity threshold is required for most of the  higher
metallicity galaxies to be excluded. Also, we notice that a comparatively high
metallicity threshold ($\Zth\sim1$) is enough to discard the artificial low-mass, high-metallicity galaxies as potential LGRB hosts. Therefore, our
conclusions are not affected by their presence in the simulation.

\begin{figure}
    \centering
    \includegraphics[width=1.0\columnwidth]{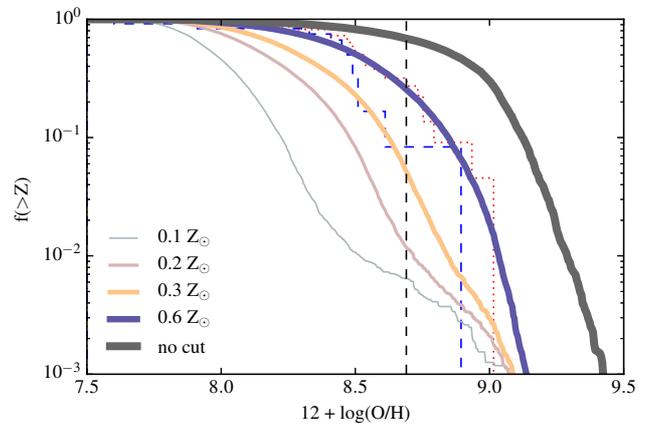}
    
    \caption{Normalized cumulative distribution of galaxies as a function of
      the metallicity of  simulated LGRB
      hosts at $z=0.7$ for models with no metallicity threshold  as
      well as those with $\Zth = 0.1, 0.2, 0.3, 0.6 \, \Zsun$. The
      black vertical  line represents the  solar oxygen abundance
      value of 8.69. The cumulative  distributions of BAT6 (blue
      dashed line) and VLT/X-Shooter sub-samples (dotted, red line) at
      $z<1$ are included for comparison.}

    \label{fig:met_dist}
\end{figure}

It should also be mentioned that the accurate determination of metallicities is a
challenging task and that several methods exists for this purpose.
Unfortunately, not all methods are reliable for all redshift and metallicity
ranges. In particular, strong-line methods are used for galaxies too faint,
distant or metal-rich for their metallicities to be measured by the preferred
electron temperature ($T_e$) method \citep{pagel_composition_1979,alloin_nitrogen_1979}. Strong-line methods have to be calibrated
against oxygen abundances previously determined either empirically, via the
$T_e$ method \citep[e.g.][]{pettini_[oiii]/[nii]_2004,pilyugin_new_2016}, or theoretically, via photoionization models \citep[e.g.][]{mcgaugh_h_1991,kobulnicky_metallicities_2004,dopita_chemical_2016}. Different
calibrations can result in large discrepancies of up to $\sim0.7$ dex \citep{moustakas_optical_2010,kewley_metallicity_2008}. Recently, \citet{curti_new_2017}
derived a new set of fully empirical calibrations for strong-line diagnostics. They
find that in the high-metallicity regime ($12+\log(O/H) > 8.2$) the
metallicities determined by the $T_e$ method are significantly lower than
those predicted by the photoionization models used by
\citet{maiolino_amaze._2008} in their calibration. This suggests that at least
part of the discrepancies found in the MZR of the Illustris simulation could
be attributed to the calibration method and that, if
\citet{maiolino_amaze._2008} metallicities were revised downwards, this would
result in a better agreement between observations and the simulation as far as metallicities are concerned.

In Fig~\ref{fig:met_dist} the normalized cumulative number distribution  of
simulated LGRB hosts  as a function of the mean metallicity of the host are
shown at $z=0.7$. The distributions are estimated for different metallicity
thresholds. It is clear that observed LGRB hosts are in disagreement with the
model with no metallicity threshold.   The best match seems to be in the range
$\Zth=0.3-0.6 \, \Zsun$. A Kolmogorov--Smirnov (K-S) test results in a
p-value\footnote{Throughout this work we use the common definition that the
p-value of the K-S test is the probability that the maximum absolute
deviation between the cumulative distributions of both samples exceeds the
observed value under the assumption that the null hypothesis, i.e. that both
samples are drawn from the same distribution, is true.} of $0.024$ and $0.059$
for $\Zth=0.3 \, \Zsun$ and $\Zth=0.6 \, \Zsun$, respectively.  For  lower
metallicities, the model with $\Zth=0.6 \, \Zsun$ indeed appears to be
consistent with observations, but the BAT6 metallicity fraction of LGRB host
with $\Zth > 8.5$ drops quickly to values more consistent with the model
corresponding to $\Zth=0.3 \, \Zsun$.

We also compare our results to the metallicity distribution derived by
\citet{kruhler_grb_2015} for their VLT/X-Shooter emission-line spectroscopy
sample of LGRB host galaxies. The advantages of this sample are the large
number of galaxies (96 galaxies) and a uniform set of measured properties
including systemic redshifts, SFRs, visual attenuations ($A_V$) and oxygen
abundances. Although their GRB selection criteria was not as carefully
designed to avoid biases as the previously mentioned samples, this sample
does include a significant number of dark and dusty LGRB hosts. In our analysis we limit the VLT/X-Shooter to $z<1$ (22 galaxies) so it can be fairly compared to both the BAT6 sample and to the
Illustris simulation at $z=0.7$. The model with $\Zth=0.6 \, \Zsun$ appears to
be a good match to these observations with a K-S test resulting in a p-value of $0.279$. The drop for high metallicity hosts is
weaker than in the case of BAT6, perhaps due to the larger sample size.
Alternatively, the larger proportion of higher metallicity hosts could be
related to the higher than average number of dusty hosts included in the
VLT/X-Shooter sample.

In both BAT6 and the VLT/X-Shooter sample, the number of hosts with a metallicity above solar is relatively
small compared to what is expected for typical star forming galaxies. From
the sub-sample of $z<1$ hosts with metallicity measurements, \citet{kruhler_grb_2015} derived a fraction of $24\pm10$ per cent of supra-solar metallicity hosts in the VLT/X-Shooter sample, while
\citet{japelj_are_2016} found a fraction of $16_{-8}^{+16}$ for the BAT6 sample
in the same redshift range. Our model with $\Zth=0.6 \, \Zsun$ presents a
similar fraction of $\sim 25$ per cent of LGRB hosts with average metallicities above solar. The fraction of supra-solar metallicity hosts falls sharply when considering stricter thresholds,
reaching less than 5 per cent for $\Zth=0.3 \, \Zsun$.

\subsection{Fundamental metallicity relation}

\begin{figure*}
    \centering
    \includegraphics[width=1.0\textwidth]{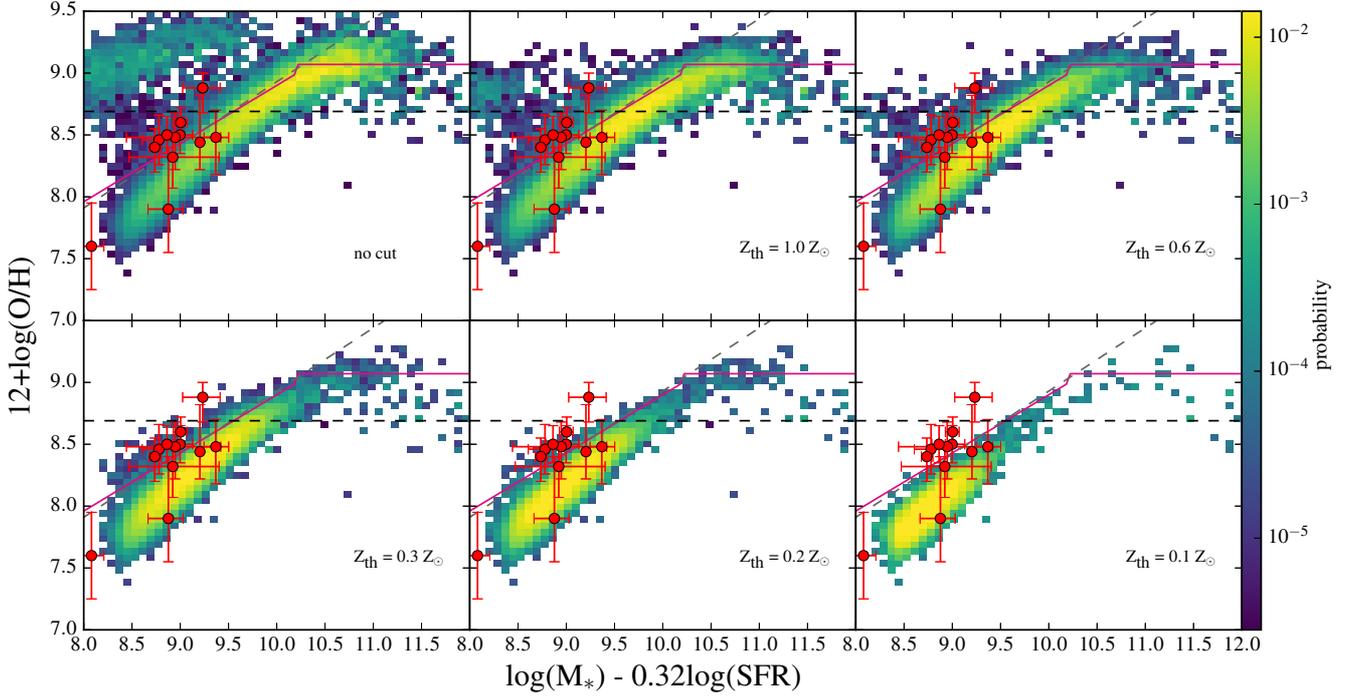}
    
    \caption{Fundamental metallicity relation of host galaxies in the
      Illustris simulation at $z=0.7$. In each panel, the galaxies in
      each bin are weighted by the probability of hosting LGRBs for
      different metallicity thresholds. For comparison to observations, we show the BAT6 sample limited to $z<1$ in
      the M08 calibration (red points). The FMR of field galaxies using the parametrised analytical models of 
    \citet{mannucci_fundamental_2010} (dashed lines) and
    \citet{mannucci_metallicity_2011} (solid lines) are also shown.}
    \label{fig:fmzr}
\end{figure*}

The fundamental metallicity relation (FMR) is a tight correlation
between stellar mass, metallicity and SFR that has been shown to be followed
by star-forming galaxies at least up to $z<2.5$ \citep{mannucci_fundamental_2010}.
\citet{mannucci_metallicity_2011} compared the FMR of field galaxies with $M_* > 10^{8.3} \,
\Msun$ to the metallicity properties of a small sample of LGRB hosts at
$z<1$. They found that LGRB host galaxies follow the FMR within errors. The same
result was found by \citet{japelj_are_2016} in the case of the $z<1$ BAT6 sample.
In Fig~\ref{fig:fmzr} we show the FMR of host galaxies in the Illustris
simulation at $z=0.7$ for different metallicity thresholds. For
comparison we also show the BAT6 sample as well as the parametrised analytical
expression of \citet{mannucci_fundamental_2010} (dashed lines) and \citet{mannucci_metallicity_2011}
(solid lines).

The results are similar to the ones found for the MZR, the simulation follows
a similar trend compared to the observed FMR, although slightly shifted
towards lower metallicity values. The model with no metallicity threshold
generates a large excess of high metallicity hosts which largely
disappears when considering a more strict $\Zth$. However, it is clear that,
in agreement with observations, the predicted LGRB hosts including a
metallicity threshold follow the general FMR but with a higher proportion of
low metallicity hosts.

It is important to point out that the host sample used by
\citet{mannucci_metallicity_2011}, while following the FMR of star-forming
galaxies, also presents a systematic offset towards lower metallicities with
respect to the MZR of field galaxies. Previous works have used the results of
\citet{mannucci_metallicity_2011} to argue against a metallicity threshold,
they claim that the apparent scarcity of high metallicity hosts could be
explained by a higher than average SFR. That conclusion was reached for
example by \citet{campisi_metallicity_2011} who used N-body simulations
combined with semi-analytical models of galaxy formation to study LGRB under
different metallicity thresholds. These results are in tension
with  both
the BAT6 sample and the Illustris simulation, which predict that
hosts follow both the MZR and FMR
simultaneously,  regardless of the metallicity threshold. It is likely that the \citet{mannucci_metallicity_2011} sample
suffered from selection biases, that have largely been corrected in the BAT6 sample.

\subsection{Star Formation Rate}

\begin{figure*}
    \centering
    \includegraphics[width=1.0\textwidth]{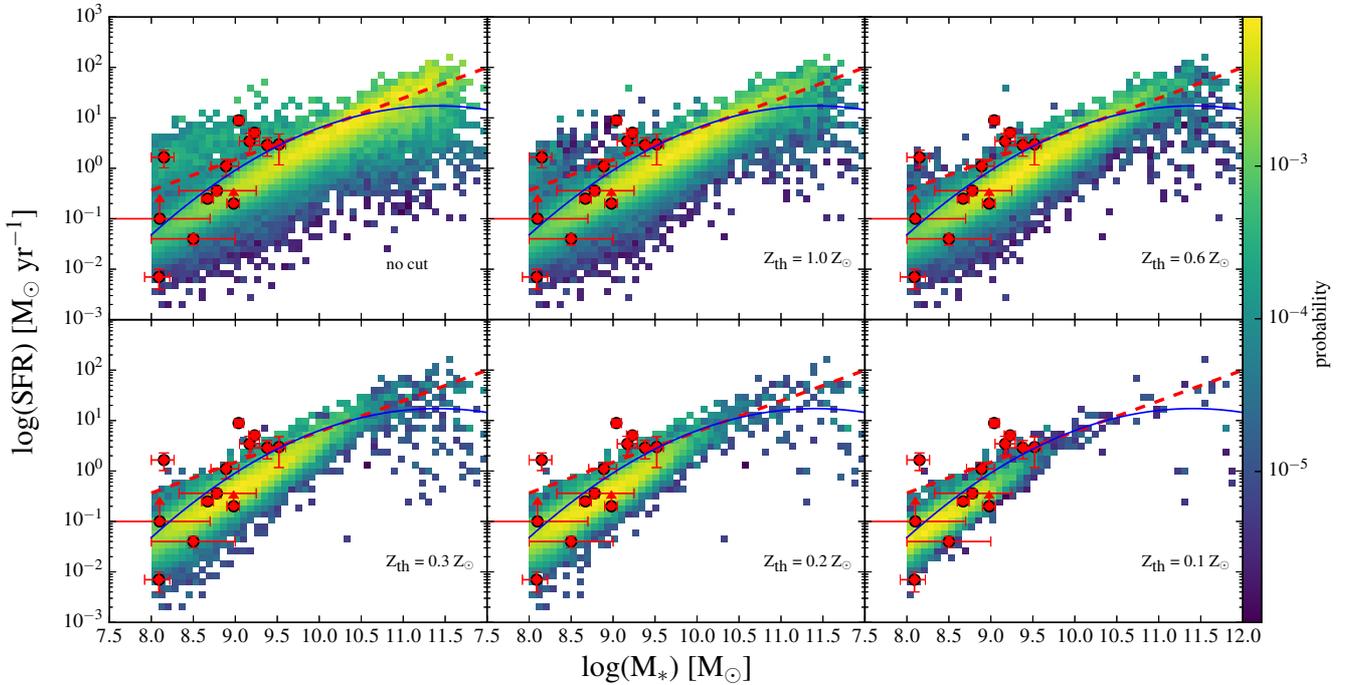}
    
    \caption{Star formation main sequence for host galaxies in the
      Illustris simulation at $z=0.7$. In each panel the galaxies in a
      given bin are weighted by the probability of hosting  LGRBs for
      different metallicity thresholds. For comparison,  we show the
      \citet{japelj_are_2016} SFMS for the $z<1$ BAT6 sample. The median
      relation derived by \citet[][dashed red
      line]{whitaker_star_2012} and \citet[][solid blue
      line]{whitaker_constraining_2014} are also displayed.}
    \label{fig:mass_sfr}
\end{figure*}

Star-forming galaxies have been shown to follow the so called star formation
main sequence (SFMS) at low \citep[$z <
1$,][]{brinchmann_physical_2004,salim_uv_2007} and high \citep[$z >1$,][]{daddi_multiwavelength_2007} redshift. The sequence is an almost linear relation between the SFR and
$M_*$, with a normalization that is observed to increase from z=0 to $2$,
around the same redshift where the $\cSFR$ also peaks.

Figure \ref{fig:mass_sfr} shows the SFMS for the simulated hosts at $z=0.7$
compared to the BAT6 sample. As pointed out by \citet{japelj_are_2016}, the SFR of the
BAT6 sample increases with stellar mass as expected. We also show for
comparison the median SFMS relation derived by \citet[][dashed red line]{whitaker_star_2012} and \citet[][solid blue line]{whitaker_constraining_2014} where they extended the
relation down to lower stellar masses ($M_* > 10^{8.4} \, \Msun$). As can be
seen, while BAT6 hosts generally follow the \citet{whitaker_star_2012} relation
with some scatter, the \citet{whitaker_constraining_2014} relation is an even better
match, in agreement with the expected lower mass of LGRB hosts.

The Illustris simulation presents a clear SFMS that is in good agreement
with observations at $z \sim 0$ and $z\sim 4$. However, the normalization of the
SFMS is lower than observations at intermediate redshifts $z\sim1$
and $z\sim 2$, as discussed in detail by \citet{genel_introducing_2014} and \citet{sparre_star_2015}. The lower normalization of the SFMS is believed to be related to a
broader peak in the Illustris $\cSFR$
\citep{vogelsberger_properties_2014} and that the
star formation and feedback processes are too closely linked to the dark
matter evolution. Reproducing the evolution of the normalization is a
challenge not only for the Illustris simulation but for theoretical
models in general \citep{dave_galaxy_2008,damen_star_2009}.

\begin{figure}
    \centering
    \includegraphics[width=1.0\columnwidth]{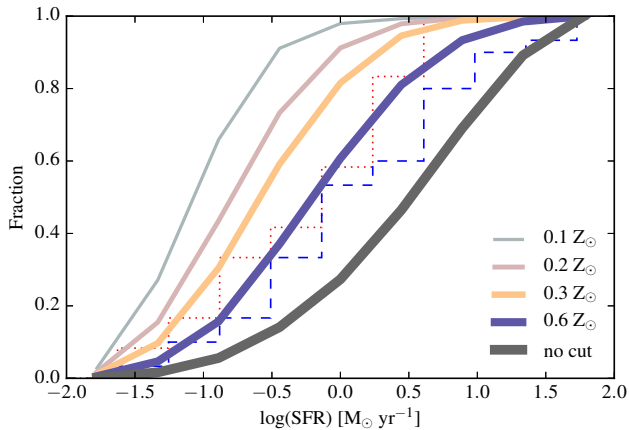}
    
    \caption{Cumulative SFR distribution of simulated LGRB hosts at $z=0.7$, considering different metallicity thresholds . The BAT6 sample
(dashed blue line) and the VLT/X-shooter sample (dotted red line) are
included for comparison. Both observational samples are limited to $z<1$.}
    \label{fig:sfr_dist}
\end{figure}

It is immediately clear from Fig. \ref{fig:mass_sfr} that in
the absence of a metallicity threshold, the simulated hosts determine  a
distribution extended towards higher SFRs than observations. We further
examine this in Fig.~\ref{fig:sfr_dist} where we plot the normalized cumulative SFR
distribution of the simulated galaxies at $z=0.7$, compared to the BAT6 sample
(dashed blue line) and the VLT/X-shooter sample (dotted red line), both
limited to $z<1$. Again, observational samples can be better matched
by a $\Zth=0.6 \, \Zsun$ model than by the no-metallicity threshold model. We use a K-S test to quantify the agreement between observations and simulations, for $\Zth=0.6 \, \Zsun$ we obtained a p-value of 0.318 (0.110) for the BAT6 (VLT/X-shooter) sample. However, we note that an increase in the normalization of the SFMS would shift all simulated models to the right, leaving the $\Zth=0.3$ model much closer to the observations.

It is important to point out that the MZR, FMR an now the SFMS of
simulated LGRB hosts favour the same $\Zth=0.6$  metallicity
threshold. Also, in every case, resolving the apparent tensions
  in the properties of simulated galaxies compared to field galaxies
  would result in a shift towards a lower metallicity threshold
  (i.e. closer to $\Zth=0.3$). All these relations are deeply correlated by star formation, chemical enrichment and feedback processes so it
is encouraging to find that the behaviour of the simulated LGRB hosts is self-consistent.

\begin{figure*}
    \centering
    \includegraphics[width=1.0\textwidth]{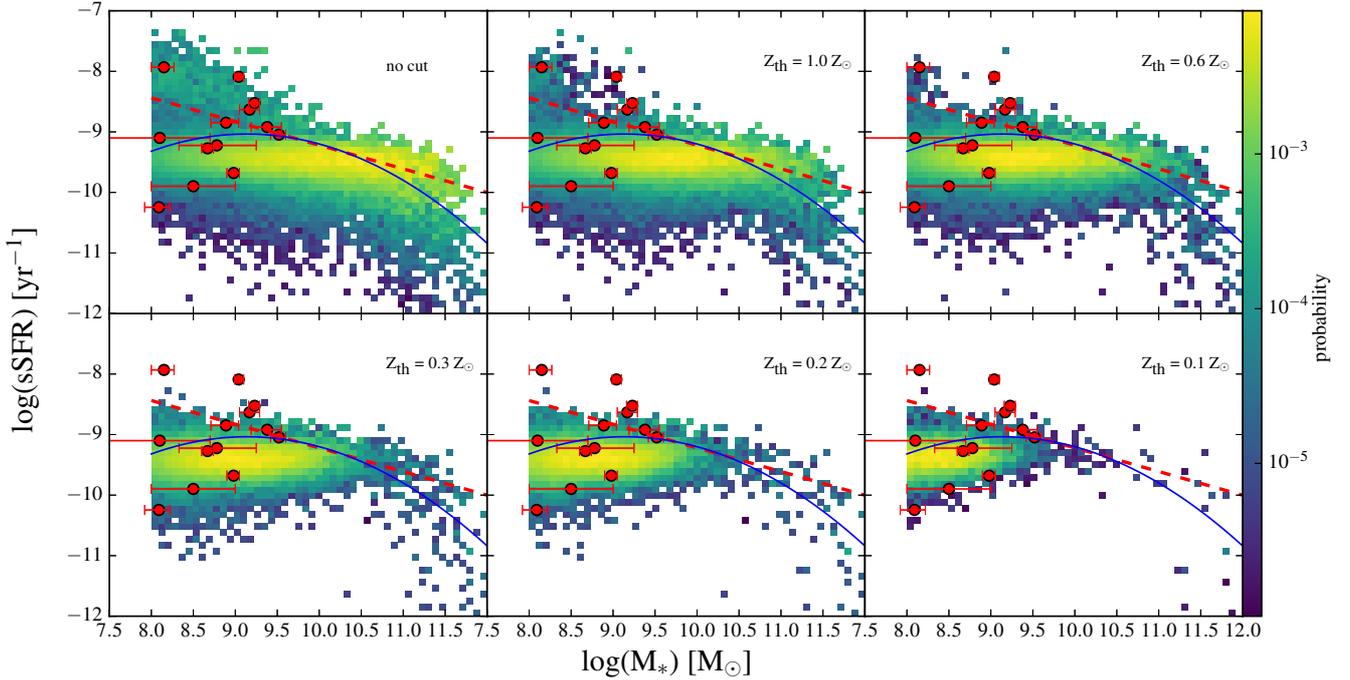}
    
    \caption{Specific star formation rate versus stellar mass for LGRB
      host galaxies in the Illustris simulation at $z=0.7$. In each
      panel galaxies in a given bins are weighted by the probability
      of hosting an LGRB according to a different metallicity
      threshold. For comparison, the \citet{japelj_are_2016} SFMS for
      the BAT6 sample is included.  The median relations derived by
      \citet[][dashed red line]{whitaker_star_2012} and
      \citet[][dashed red line]{whitaker_constraining_2014} are also shown.}
    \label{fig:mass_ssfr}
\end{figure*}

The sSFR in Illustris inherits the normalization problems that originate
in the SFMS. Additionally, at a given redshift, it becomes approximately
independent of mass for $M_* < 10^{10.5} \, \Msun$ which is at
odds with observations by \citet{behroozi_average_2013} and \citet{whitaker_star_2012} (Fig.~\ref{fig:mass_ssfr}, first panel). The relation of Whitaker et
al. (2014), on the other hand, does show a turn around of the sSFR at lower
stellar masses which is more similar to the behaviour of the simulation. \citet{japelj_are_2016} studied the sSFR of the BAT6 sample and found that a significant
fraction ($27_{-9}^{+15}$ per cent) of hosts could be classified as starbursts
according to their criteria. In Fig.~\ref{fig:mass_ssfr} the LGRBs hosts
appear to be equally scattered around the intrinsic sSFR of the simulation, for
all metallicity thresholds considered. This suggests no preference
for star-bursting LGRBs hosts in the Illustris simulation. This lack of starbursting hosts could be related to results by \citet{sparre_star_2015}, who found a paucity of strong starbursts
in the Illustris simulation presumably due to limited resolution that
prevents tidal torques which drive starbursts during
mergers. In this case, it is a caveat of the Illustris simulations and
not a problem of the the LGRB model adopted in this work.

\subsubsection{SFR redshift evolution}

\begin{figure}
    \centering
    \includegraphics[width=1.0\columnwidth]{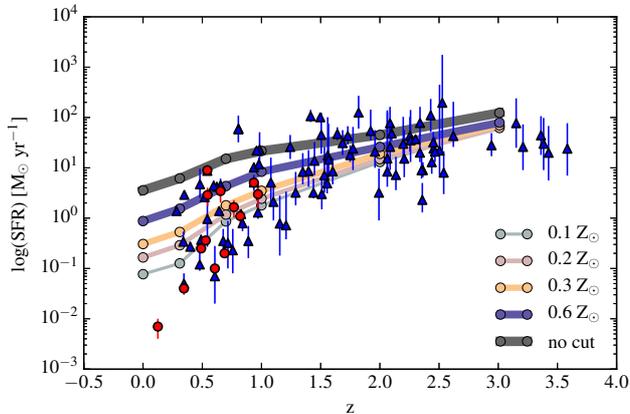}
    
    \caption{Evolution of the SFR of LGRB host galaxies as a
      function of redshift in the Illustris simulation for
      models with different $\Zth$ (solid lines). The blue triangles
      represent the VLT/X-shooter sample. Red dots represent the BAT6
      sample for $z<1$.}
    \label{fig:sfr_in_z}
\end{figure}

\citet{kruhler_grb_2015} found a clear evolution in the SFR of LGRB hosts with
redshift. The observed median SFR of LGRB host increases from $\sim0.6
\, \Msun \, \yr^{-1}$ at $z\sim0.5$ by a factor of 25 to $\sim15 \, \Msun \,
\yr^{-1}$ at $z\sim2$. For higher redshift only a weak evolution is
observed, with the SFR levelling off at around $\sim20 \, \Msun \, \yr^{-1}$ at
$z\sim2$. In Fig.~\ref{fig:sfr_in_z} we explore the evolution in SFR of
Illustris LGRB hosts. To do so, we plot the mean SFR of LGRB hosts, for each
metallicity threshold, as a function of redshift. 

In order to properly compute the mean evolution of the SFR, a
redshift dependent lower SFR limit that matches the average SFR sensitivity of
the VLT/X-shooter sample \citep[see figure 13 in][]{kruhler_grb_2015} was
considered. Compared to the VLT/X-shooter sample (blue triangles), all models
present the same trend of increasing SFR with redshift up to $z\sim3$, but only
models with a metallicity threshold can match the low levels of SFR at low
redshift. As shown in Fig.~\ref{fig:sfr_in_z},  the BAT6 sample for
$z<1$ (red dots)  is also
compatible with the VLT/X-shooter sample,  reinforcing the previous
conclusion.

\section{Internal metallicity dispersion}

As pointed out in section \ref{sec:mzr}, despite the paucity of high
metallicity hosts, there is observational evidence of highly chemically
enriched LGRB hosts, with metallicities larger than solar
\citep{graham_high_2015}. This is also reproduced by the Illustris
simulation, even when imposing a metallicity threshold for LGRB progenitors.
This is possible because the  ISM in a galaxy is not chemically homogeneous.
The internal dispersion and variation of chemical abundances of the ISM are
observationally proven in our Galaxy and in other galaxies \citep[e.g.][]{afflerbach_galactic_1997,smartt_chemical_2001,sanders_metallicity_2012}.  Hence,
the metallicity of LGRB progenitor sites might differ  from the mean
metallicity of the host galaxy.

Some tentative evidences of such effect has been reported for a small number
of local galaxies, for which the LGRB region could be spatially-resolved
\citep{christensen_ifu_2008,thone_spatially_2008,levesque_metallicity_2011}.  They found that LGRB sites may
indeed have slightly lower metallicities than the host average. However, more
detailed studies remain challenging as the number of spatially-resolved hosts
is very limited.

Metallicity gradients may also play an important role. \citet{artale_chemical_2011}
found that scenarios favouring low-metallicity progenitors tend to produce
LGRBs further out from the central regions than those allowing high
metallicity progenitors at low $z$. However, some metallicity scatter is expected at each
galactocentric radius, this scatter can be equal or even greater than the
radial variation \citep{sanders_metallicity_2012}. We therefore explore the role of
metallicity dispersion ($\sigma_{\rm Z}$) in producing high metallicity hosts.

\begin{table*}
\caption{ Best-fit parameter values for the $\sigma_{\rm Z}-{\rm Z}$ correlations in Fig.~\ref{fig:met_disp} corresponding to two possible parametrizations, a linear function given by $\sigma_{\rm Z}= a {\rm Z} + b$ and a logistic curve give by $\sigma_{\rm Z}= c / (1+\exp(-k({\rm Z}-d)))$}
\label{tab:met_disp_par}
\centering
\begin{tabular}{lccccc}
\hline
$Z_{\rm th}$ & $a$ & $b$ & $c$ & $d$ & $k$ \\
\hline
no cut & $+0.004\pm0.008$ & $+0.15\pm0.07$ & $+0.19\pm0.001$ & $+8.05\pm0.05$ & $+9.01\pm2.64$ \\
1.0 & $+0.031\pm0.007$ & $-0.08\pm0.07$ & $+0.20\pm0.002$ & $+8.04\pm0.04$ & $+7.32\pm1.49$ \\
0.6 & $+0.084\pm0.008$ & $-0.54\pm0.07$ & $+0.21\pm0.005$ & $+8.02\pm0.04$ & $+4.33\pm0.79$ \\
0.3 & $+0.191\pm0.010$ & $-1.43\pm0.09$ & $+0.27\pm0.022$ & $+8.17\pm0.05$ & $+3.43\pm0.60$ \\
0.2 & $+0.257\pm0.012$ & $-1.96\pm0.1$ & $+0.3\pm0.034$ & $+8.20\pm0.07$ & $+3.79\pm0.60$ \\
0.1 & $+0.362\pm0.021$ & $-2.76\pm0.17$ & $+0.37\pm0.062$ & $+8.12\pm0.08$ & $+4.48\pm0.71$ \\
\hline
\end{tabular}
\end{table*}

\begin{figure*}
    \centering
    \includegraphics[width=1.0\textwidth]{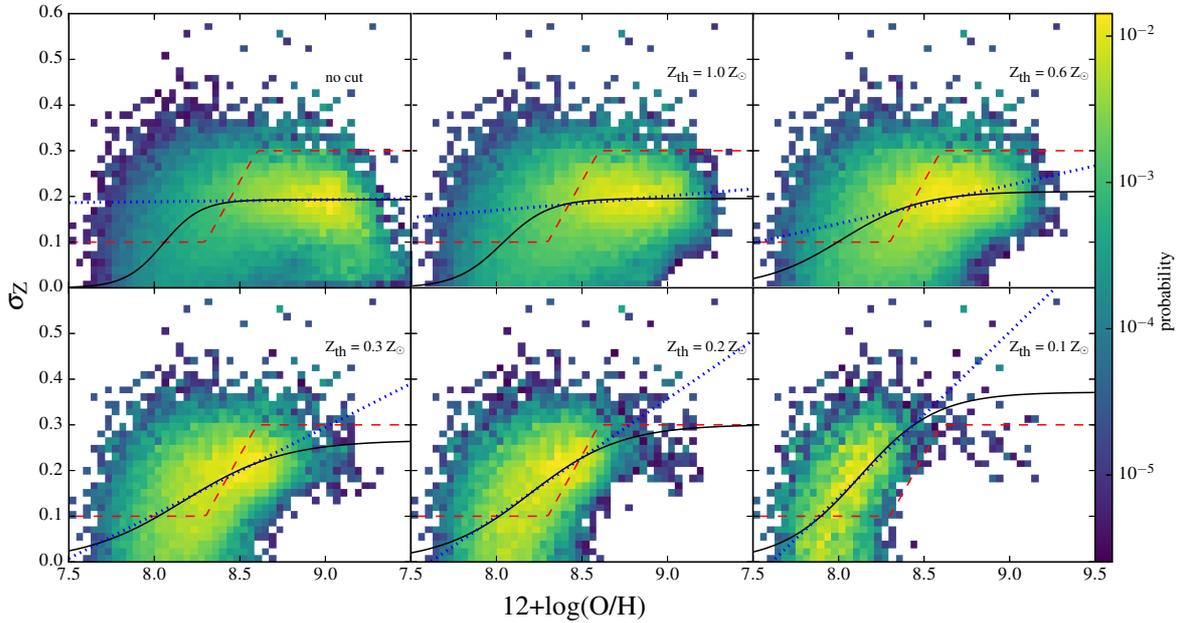}
    
    \caption{Internal metallicity dispersion of the host LGRBs galaxies in the Illustris simulation at $z=0.3$ as a function of average galaxy metallicity. In each panel the bins are weighted by the probability of hosting an LGRB according to different metallicity thresholds. Solid (dotted) lines represent fits to the probability weighted maps assuming a logistic (linear) model. The red dashed line represents the internal metallicity dispersion model proposed by \citet{niino_redshifted_2016}. }
    \label{fig:met_disp}
\end{figure*}

\citet{niino_redshifted_2016} studied available LGRB hosts with $z<0.41$. They
compared the metallicity distribution to model predictions taking into account
the spatial variation of metallicities among star forming regions within a
galaxy. They found that models in which only low-metallicity stars
(12+log(O/H)$<8.2$) can trigger LGRBs reproduce the observed metallicity
distribution. This is close to the value favoured by the Illustris simulation.

For our simulated LGRB host galaxies, it is possible to analyse the
internal metallicity distribution in relation to the metallicity of
the progenitor star.
In Fig.~\ref{fig:met_disp} we show the internal metallicity dispersion as
a function of the average host metallicity at $z=0.3$. We choose to
focus in a redshift range similar to that of the  LGRB hosts
studied by \citet{niino_redshifted_2016}. As can be seen, in the absence of a
metallicity threshold, there is practically no correlation between the two quantities.
But one does gradually appears when considering lower $\Zth$. We have
quantified the correlation using a linear relation as well as a
logistic curve. The latter  matches the sigmoid appearance of the
simulated results. The fitting parameters  are presented in Table \ref{tab:met_disp_par}.

At $\Zth=0.2-0.3 \Zsun$ the simulated metallicity dispersion largely agrees
with that proposed by \citet{niino_redshifted_2016}. It is important to
note that the correlation between metallicity and internal metallicity
dispersion arises as a consequence of the metallicity threshold. This implies
that, under the assumption of a stringent metallicity threshold ($\Zth = 0.1 -
0.3 \Zsun$), the simulation predicts that high metallicity hosts, which
simultaneously have a high internal metallicity dispersion, are likely to be
found.

In the case of models with $\Zth=0.2-0.3 \Zsun$, the difference in $\sigma_{\rm
Z}$ between low-metallicity hosts and high metallicity is not very large ($\sim
0.2 \, {\rm dex}$). This is in agreement with observations that show  only a
small decrease in the  metal content of  LGRB sites, compared to the average host
metallicity. Also, by limiting the metallicity of the stellar progenitor instead of 
the metallicity of the host galaxy, our models favour a lower
metallicity threshold in agreement with
\citet{niino_redshifted_2016}. Other studies \citep{wolf_metallicity_2007,kocevski_modeling_2009,perley_swift_2016-1,japelj_are_2016} {found instead a} threshold closer to solar metallicity for efficient LGRB production.

\section{Mass distribution}

\begin{figure}
    \centering
    \includegraphics[width=1.0\columnwidth]{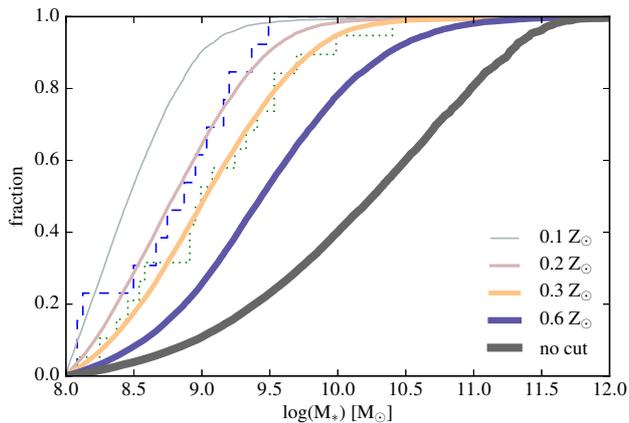}
    
    \caption{Cumulative mass distribution of simulated LGRB hosts at $z=0.7$ for models
    with no metallicity threshold, as well as $\Zth = 0.1, 0.2, 0.3, 0.6 \,
    \Zsun$. The blue dashed line represents the mass distribution of the BAT6
    sample, while the green dotted line represents the SHOALS sample. Both
    observational samples are restricted to $z<1$}

    \label{fig:mass_dist}
\end{figure}

Beside low-metallicity and low-SFR, another feature of LGRB hosts is their low
stellar mass at $z<1$. \citet{vergani_are_2015} found that, in that redshift
range, the BAT6 sample mass distribution was significantly shifted towards
lower masses compared to galaxies in the UltraVISTA survey. Similarly,
\citet{perley_connecting_2015} found a preference of LGRBs to occur in low stellar mass galaxies for a complete sample of radio-observed LGRB hosts.

\citet{perley_swift_2016-1} studied the rest-frame NIR luminosities and stellar
masses of the SHOALS sample. They found a rapid increase in the NIR host
luminosity between $z\sim0.5$ and $z\sim1.5$, and little variation for higher
redshifts. In Fig. \ref{fig:mass_dist} we plot the cumulative stellar mass
distributions of the BAT6 and SHOALS sample at $z<1$ together with the
distribution of our simulated hosts considering different metallicity
thresholds. Our results strongly discard that the observational samples could
be drawn from the no metallicity threshold model. Compared to SHOALS, the BAT6
sample presents slightly lower stellar mass hosts and is better described by a
$\Zth=0.2 \Zsun$ model (K-S test p-value=0.491), while the SHOALS sample is better described by a
$\Zth=0.3 \Zsun$ model (K-S test p-value=0.606).

We remark that the metallicity threshold favoured by the simulation
after analysing the stellar mass distribution ($\Zth \sim0.3 \, \Zsun$)
is lower than what we found for the MZR, FMR and SFMS ($\Zth \sim0.6 \,
\Zsun$). Unlike these latter relations, for which Illustris presents
some departure from observed field galaxies, the stellar mass
distribution of simulated galaxies is a close match to observations at all
redshifts \citep{genel_introducing_2014}. Therefore, we point out that
if the tensions in MZR and SFR between simulation and observations were
resolved, the preferred threshold would be more in line to that found in this
section ($\Zth \sim0.3 \, \Zsun$).

\citet{vergani_are_2015} also studied the effect of a metallicity bias on  the
stellar mass distribution of LGRB hosts. To do so, they applied a method
similar to \citet{campisi_properties_2009}, using a galaxy catalogue
constructed by combining high-resolution N-body simulations with a semi-analytic model of galaxy formation. Their results favoured a metallicity
threshold $\Zth = 0.3-0.5 \Zsun$, which is agreement with the results found in
previous sections, but is higher than what we find studying exclusively the stellar mass distribution.

Similarly, \citet{perley_swift_2016-1} proposed a simple model in which the LGRB
efficiency is constant at low metallicity but falls sharply above a maximum
metallicity threshold to explain the decrease of host stellar masses in the
SHOAL sample at $z<1.5$. They found a close to solar metallicity threshold,
much higher than the value we find. Neither of the approaches taken by
\citet{vergani_are_2015} and \citet{perley_swift_2016-1} took into
account the internal metallicity dispersion of the galaxy, which would
result in a lowering of the metallicity threshold in {agreement} with the
results we show in the previous sections. It is therefore encouraging than further improvements in sub-grid physical models are likely to result in a better match of the present LGRB model to observations

\begin{figure}
    \centering
    \includegraphics[width=1\columnwidth]{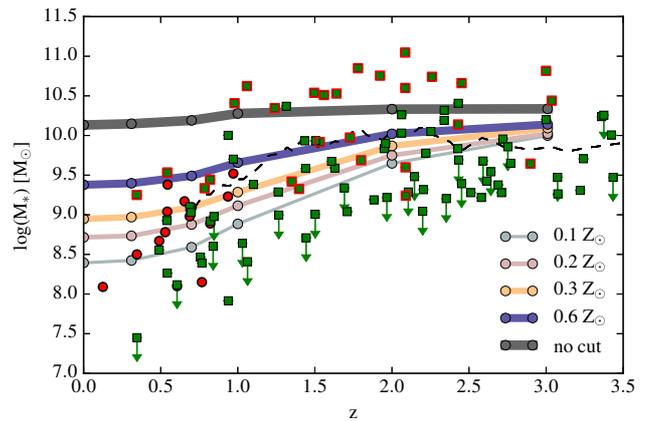}
    
    \caption{Evolution of the mean stellar mass of LGRB host as a function of redshift in the Illustris simulation under models with different $\Zth$ (solid lines). The red dots represent the $z<1$ BAT6 sample. Green square represent the SHOALS sample. SHOALS hosts that present any evidence of dust obscuration are marked with an orange border. The dashed line represents the running mean of the SHOALS sample.}
    \label{fig:mass_in_z}
\end{figure}

We test whether the Illustris simulation is capable of reproducing the
evolution of hosts stellar mass with redshift by computing the mean stellar
mass as a function of redshift, weighted by the probability of hosting an LGRB
assuming different metallicity thresholds. To compute the mean
mass we assumed a redshift dependant lower mass limit that is compatible
with the $m=25$ NIR apparent magnitude detection limit of the SHOALS
sample. The results are shown in Fig.~\ref{fig:mass_in_z} where we also
display the SHOALS and BAT6 samples. We find that all models predict a
mean host mass of $\sim10^{10} \, \Msun$ at $z=3$. However, only models with a
metallicity threshold can accurately reproduce the downsizing of LGRB
host mass for lower redshift. Metallicity thresholds between $\Zth=0.2 -
0.6 \, \Zsun$ better reproduce the running mean of the SHOALS sample at lower
redshift.

\section{Conclusions}

In this paper, we study the properties of LGRB host galaxies, comparing the
latest observational hosts samples with a catalogue of simulated galaxies
constructed from Illustris, a state-of-the-art hydrodynamical cosmological
simulation. This kind of simulations, allows us to consider relevant
processes for galaxy evolution, such as star formation and chemical enrichment,
in a self consistent way. As well as to differentiate the properties of
potential LGRB progenitors from the average properties of their host galaxies.

The aim of our investigation is to determine the validity of the metallicity
dependant model. To do so we determine the probability of galaxies in the
simulation to be considered hosting LGRBs under the assumption of the
existence of selected metallicity thresholds: $\Zth = 0.1, 0.2, 0.3, 0.6, 1.0
\, \Zsun$, as well as the null hypothesis of no metallicity dependence.

Our analysis shows that at $z<1$, if no metallicity threshold is assumed, the
MZR of simulated hosts is very different compared to the BAT6 sample, in
particular, there is a large excess of simulated high-metallicity, high stellar mass
hosts. We conclude from a more detailed study of the metallicity distribution,
that BAT6 and VLT/X-Shooter observations are more consistent with a
metallicity threshold in the range $\Zth = [0.3 - 0.6] \, \Zsun$. We also find
that the effect of the metallicity threshold in the MZR of hosts is basically
to greatly reduce the probability of high-metallicity galaxies to hosts LGRBs.
We find that lower metallicity hosts follow the intrinsic MZR of field
galaxies in the simulation, with no apparent shift towards lower
metallicities. This is in contrast to previous results based in semi-analytic
simulations of galaxy formation \citep{campisi_metallicity_2011}.

The fraction of simulated LGRB hosts with metallicities above solar is $\sim
25$ percent under the assumption of models with $\Zth = 0.6 \, \Zsun$. This
value is comparable to the fractions of high metallicity hosts found in
observational samples, in particular the $24\pm10$ per cent of supra-solar
metallicity hosts in the VLT/X-Shooter found by \citet{kruhler_grb_2015} and
the $16_{-8}^{+16}$ percent found by \citet{japelj_are_2016} for the BAT6
sample.

Metallicity dispersion plays an important role in the presence of highly
enriched LGRB hosts.  We find that, if a  $\Zth \sim 0.6 \, \Zsun$ threshold
is assumed, then high metallicity hosts also exhibit a larger internal
metallicity dispersion. We have quantified the correlation between internal
metallicity dispersion and averaged host metallicity using fits to a linear
and logistic models. We also notice that the internal metallicity dispersion of
hosts assuming $\Zth = 0.3 - 0.2 \, \Zsun$ predicted by Illustris roughly
matches the model proposed by \cite{niino_redshifted_2016} to explained the
properties of $z<0.4$ LGRB hosts.

The FMR of simulated hosts presents a similar behaviour than the MZR. Host
galaxies follow the intrinsic FMR of simulated galaxies, with the effect of a
metallicity threshold being to make lower metallicity galaxies the more likely
hosts. Again, we find a better match to observations for thresholds in the
$\Zth = [0.3 - 0.6] \, \Zsun$ range. Therefore, we conclude that at low
redshift simulated LGRB hosts follow both the FMR and MZR simultaneously, but
with a paucity of high-metallicity hosts, in accordance with observational
results by  \citet[e.g.][]{japelj_are_2016}.

From studying the SFMS of LGRB hosts we find that in the absence of a
metallicity threshold, the star formation of LGRB hosts at $z\sim1$ extends
towards much higher SFRs than in the BAT6 and VLT/X-shooter samples.
Observations are better matched by a $\Zth = 0.6 \, \Zsun$ threshold, but given
that the SFMS of the Illustris simulation presents generally a lower
normalization than observed galaxies, a lower threshold $\Zth
= 0.3 \, \Zsun$ is possible. We also find that the Illustris simulation reproduces the
evolution of the SFR of LGRB hosts with redshift in the VLT/X-shooter sample
when a metallicity threshold is assumed. Under all conditions, simulated hosts
increase their average SFR with redshift up to $z=3$, but only models with a
strong metallicity threshold reproduce the lower SFRs of low redshift hosts.

Finally, the stellar mass distribution of LGRB hosts at $z<1$ cannot be
reproduced by Illustris without a metallicity threshold. The stellar
mass distribution of the SHOALS sample is best described by a metallicity
threshold $\Zth = 0.3 \, \Zsun$, while the BAT6 sample is better described by
a lower value of $\Zth = 0.2 \, \Zsun$. We notice that this thresholds are much
lower than the close to solar metallicity values estimated by
\cite{japelj_are_2016} for the BAT6 sample and \cite{perley_swift_2016-1} for
the SHOALS sample. Thanks to the nature of the hydrodynamical simulation, we
are able to easily differentiate between the metallicity of LGRB progenitors
and the metallicity of LGRB hosts and take into account the internal
metallicity dispersion of galaxies seamlessly, this results in lower
metallicity threshold that is closer to what is expected under the collapsar
progenitor model. We also find that models with $\Zth = 0.3 - 0.6\, \Zsun$,
correctly predict the increase in stellar mass with redshift of LGRB host
present in the SHOALS sample.

Illustris provides us with the largest sample of simulated well-resolved
LGRB hosts so far analysed . Certainly, there is still room for improvement of
subgrid models, specially those related to feedback and chemical enrichment,
but globally, the simulation is able to reproduce a large number of key LGRB
host properties, their MZR, FMR, SFR and stellar mass. The internal
metallicity dispersion of galaxies seems to be crucial to understand the origin
of LGRBs and their relation with their host galaxies. All the results support
the existence of a metallicity threshold in the order $\Zth = 0.3 - 0.6\,
\Zsun$, which is lower than some of the estimations reported by previous
works.

\section*{Acknowledgements}

The authors acknowledge the grants PICT 2011-0959 from Argentinian ANPCyT, and
PIP 2012-0396 from Argentinian CONICET and Fondecyt Regular 115033, Southern
Astro-physics Network Redes Conicyt 150078 and proyecto interno MUN UNAB
2015. This research made use of Astropy
\citep{astropy_collaboration_astropy:_2013}, numpy \citep{walt_numpy_2011},
and matplotlib \citep{hunter_matplotlib:_2007}.

\bibliographystyle{mnras}
\bibliography{paperII}

\bsp    
\label{lastpage}
\end{document}